\documentclass{article}

\usepackage{PRIMEarxiv}

\usepackage[utf8]{inputenc} % allow utf-8 input
\usepackage[T1]{fontenc}    % use 8-bit T1 fonts
\usepackage{hyperref}       % hyperlinks
\usepackage{url}            % simple URL typesetting
\usepackage{booktabs}       % professional-quality tables
\usepackage{amsfonts}       % blackboard math symbols
\usepackage{nicefrac}       % compact symbols for 1/2, etc.
\usepackage{microtype}      % microtypography
\usepackage{lipsum}
\usepackage{fancyhdr}       % header
\usepackage{graphicx}       % graphics
\graphicspath{{figures/}}     % organize your images and other figures under media/ folder
\usepackage{pdfpages}
\usepackage{graphicx}
\usepackage{subcaption}
\usepackage{epsfig}
\usepackage{float}

\title{Urban Sound Propagation: a Benchmark for 1-Step Generative Modeling of Complex Physical Systems
}

\author{
  Martin Spitznagel, Janis Keuper \\
  Institute for Machine Learning and Analytics (IMLA)  \\
  Offenburg University, Germany \\
  \texttt{martin.spitznagel@hs-offenburg.de}\\
  \texttt{keuper@imla.ai}\\
  \\
  \textbf{Reviewed on OpenReview:} \url{https://openreview.net/forum?id=1vCAi53AVj}
}

\begin{document}
\maketitle

\begin{abstract}
Data-driven modeling of complex physical systems is receiving a growing amount of attention in the simulation and machine learning communities. Since most physical simulations are based on compute-intensive, iterative implementations of differential equation systems, a (partial) replacement with learned, 1-step inference models has the potential for significant speedups in a wide range of application areas. In this context, we present a novel benchmark for the evaluation of 1-step generative learning models in terms of speed and physical correctness.\\
\noindent Our \textit{Urban Sound Propagation} benchmark is based on the physically complex and practically relevant, yet intuitively easy to grasp task of modeling the 2d propagation of waves from a sound source in an urban environment. We provide a dataset with $100k$ samples, where each sample consists of pairs of real 2d  building maps drawn from \textit{OpenStreetmap}, a parameterized sound source, and a simulated \textit{ ground truth} sound propagation for the given scene. The dataset provides four different simulation tasks with increasing complexity regarding reflection, diffraction and source variance. A first baseline evaluation of common generative \textit{U-Net, GAN} and \textit{Diffusion} models shows, that while these models are very well capable of modeling sound propagations in simple cases, the approximation of sub-systems represented by higher order equations systematically fails. \\
\noindent Information about the dataset, download instructions and source codes are
provided on our website: \url{https://www.urban-sound-data.org}.
\end{abstract}


\section{Introduction}

\noindent Critical for urban planning and noise pollution management, traditional sound mapping methods are resource-heavy~\cite{roadNoise}. Our approach uses a dataset from OpenStreetMap, annotated with simulated sound maps to mirror diverse urban scenarios~\cite{ijerph14101139}. The research evaluates the effectiveness of U-Net, Generative Adversarial Networks (GAN)~\cite{goodfellow2014generative}, and Denoising Diffusion Probabilistic Models~\cite{ho2020denoising} in adhering to the complexities of urban soundscapes. We specifically investigate these models' capabilities in capturing sound reflections and diffractions.

\noindent In the realm of generative model development for physical phenomena, the availability and specificity of datasets play a crucial role. While datasets are catering to a variety of environmental and structural scenarios, there remains a noticeable gap in resources specifically designed for urban sound propagation. This gap is particularly significant given the complex interplay of variables in urban environments that affect sound dynamics, such as building layouts, material properties, and ambient conditions.

\noindent Our research fills this void by introducing a novel dataset for studying urban sound propagation. This dataset is distinct in its focus on the intricate patterns and behaviors of sound waves as they navigate through urban landscapes. The introduction of this dataset marks a pivotal advancement in the field. It provides a much-needed resource for precise simulation and analysis of urban soundscapes and sets a new standard for future explorations in this area. By offering a dataset specifically focused on urban sound dynamics, we aim to catalyze further research and innovation in sound mapping using advanced machine learning techniques, thereby contributing to the evolution of urban planning and noise mitigation strategies.

\subsection{Related Work}

The integration of physical principles within generative models represents an active and burgeoning area of research, particularly in fields such as image and sound processing. Models like PUGAN~\cite{PUGAN} and FEM-GAN~\cite{FEM-GAN} have showcased the potential of combining GANs with physical modeling to enhance performance in environments with complex physical laws. Likewise, progress in fluid dynamics or structural system identification via the PG-GAN approach has highlighted the improvements in both efficiency and precision achievable when generative models incorporate physics-based loss functions and simulations~\cite{Kim2018Deep, YU2024122339}.

\noindent These studies collectively underscore a crucial trend: the integration of physical laws into the training of generative models not only enhances the fidelity and reliability of these models but also significantly improves their performance in tasks involving complex physical phenomena~\cite{PhysDiff, PhysicsGuided, PhysicallyGrounded}. While significant advancements have been made in the realm of generative models across diverse disciplines, the domain of urban sound propagation presents a unique set of challenges that remain largely unaddressed. Recognizing this deficiency, our research underscores the necessity of incorporating physics-guided principles into generative modeling. By doing so, we aim to establish a new baseline for urban sound propagation research, marking an essential first step towards 1-Step generative modeling of complex physical systems.

\section{Physics of Sound Propagation}

\noindent Mathematically, the propagation of sound over time is described via partial differential wave equations. Due to space constrains and the practical nature of our problem setting, we will neglect the derivation from continuous wave equations and directly focus an the discrete and iterative implementations of sound propagation which have been applied for our \textit{ground-truth} simulations.\\
Following \cite{vorlander2007auralization}, for a discrete set of receivers $R$, the amplitude  $L^{j}_{{R_k}}$ of receiver $R_k$ at frequency $j$ is computed via iterative differences:
\begin{eqnarray}
\label{eq:noise}
L^{j}_{{R_k}} &=& L^{j}_{W} - A_{div_{R_{k}}} - A^{j}_{atm_{R_k}} - A^{j}_{dif_{R_{k}}} - A^{j}_{grd_{R_{k}}} \textrm{\quad, where}\\
& L^{j}_{W} &\textrm{models the source,}\\
& A_{div_{R_{k}}} &\textrm{captures the geometrical spreading,}\\
& A^{j}_{atm_{R_k}} &\textrm{represents the atmospheric absorption,}\\
& A^{j}_{dif_{R_{k}}} &\textrm{models diffraction,}\\
& A^{j}_{grd_{R_{k}}} &\textrm{the ground effect - which is neglected in our study.}
\end{eqnarray}

%In this formula, $A_{div_{R_{k}}}$ is the geometrical spreading, $A^{j}_{atm_{R_k}}$ is atmospheric absorption, and $A^{j}_{dif_{R_{k}}}$ represents diffraction. The ground effect, $A^{j}_{grd_{R_{k}}}$, was omitted for simplicity.

\noindent Additionally, the model accounts for reflections by adjusting the power level $L^{j}_{W}$ based on the absorption coefficient $\alpha_{vert}$ of the surfaces involved. This adjustment is performed using the equation
\begin{equation}
\label{eq:reflection}
L^{(n_{ref})}_{W} =  L^{(n_{ref} - 1)}_{{W}} + n_{ref} \times 10 \log_{10} (1 -{\alpha}_{vert})
\end{equation}
where $n_{ref}$ indicates the number of reflections considered. Specular reflections are modeled using the image receiver method, which provides a computationally efficient way to account for the angle of incidence being equal to the angle of reflection~\cite{vorlander2007auralization}. The sound level at each receiver reflects the cumulative effect of direct, diffracted, and reflected sound paths.

\noindent As the number of reflections increases, the complexity of calculating the power level $L^{j}_{W}$ also increases. This is because each reflection path introduces additional calculations that are not strictly linear due to the logarithmic nature of the decibel scale and the multiplicative effect of each reflection's absorption. %The accumulation of these effects transforms what might be assumed as a static power level into a non-linear function, heavily influenced by the environmental geometry and surface characteristics. The interplay of these factors results in a more complex and nuanced simulation of sound propagation, reflecting the intricate ways in which sound interacts with its surroundings.
\section{Dataset Creation and Properties}

The proposed dataset has been generated from $25k$ real geolocations, spread across 10 different pre-selected cities. For each of these cities, our dataset provides 2,500 samples from distinct locations, thereby offering a wide range of urban environments for our analysis. The associated image samples originate from pre-processed satellite images that capture the urban landscape within a rectangular 500m² area, depicting buildings as black pixels, while open spaces are represented by white pixels. All of this geodata has been collected from open sources via the \textit{Overpass API}~\footnote{OSM-data for this study was collected using the Overpass API, available at \url{http://overpass-api.de/api/map}.} and has been processed with \textit{GeoPandas}~\cite{kelsey_jordahl_2020_3946761} to ensure a consistent coverage of 500m² per sample.
Further, we applied a location selection heuristic to steer the sample positioning relative to building structures and required each site to be encircled by a minimum of 10 buildings within a 200-meter radius. Additionally, a constraint of a 50-meter minimum distance from any building to the sample location was imposed to model realistic urban scenarios.

\noindent\textbf{Ground-Truth Simulation.} We applied the open source \textit{NoiseModelling v4} simulation framework~\cite{noisemodelling_framework}, which fulfills the \textit{CNOSSOS-EU} standard for noise emission~\cite{CNOSSOS-EU} in order to generate \textit{ground-truth} propagation maps for the collected samples according to eq. \ref{eq:noise}. We extended the existing regular 5m simulation grid by placing additional sound receivers at every 5 meters along the edges and at corners of buildings, while systematically removing any receivers positioned inside the buildings. After the simulation step, we linearly interpolated the receiver values onto a 512x512 or 256x256 pixel maps, where each pixel indicates the decibel level at that specific location. The decibel values are normalized to a grayscale range, mapping a 0-100 dB range to 0-255 grayscale values. Fig. \ref{fig:overview} gives an overview of the full process. More details on our scalable implementation of the generation pipeline are given in Appendix \ref{app:pipeline}. 

\begin{figure}[h]
\centering
\begin{subfigure}[b]{0.225\linewidth}
    \includegraphics[width=\linewidth]{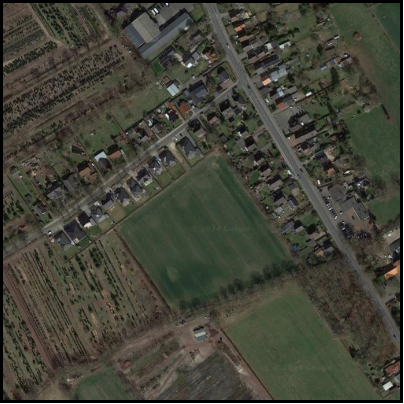}
    \caption{Sattelite image}
    \label{fig:imageA}
\end{subfigure}
\hfill
\begin{subfigure}[b]{0.225\linewidth}
    \includegraphics[width=\linewidth]{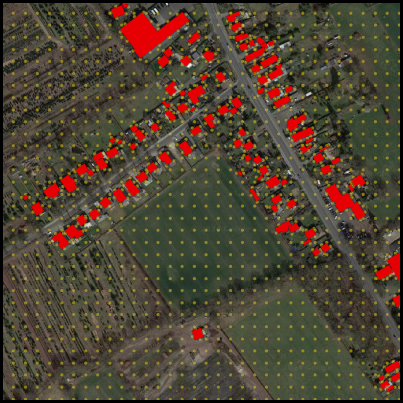}
    \caption{Receiver grid}
    \label{fig:imageB}
\end{subfigure}
\hfill
\begin{subfigure}[b]{0.225\linewidth}
    \includegraphics[width=\linewidth]{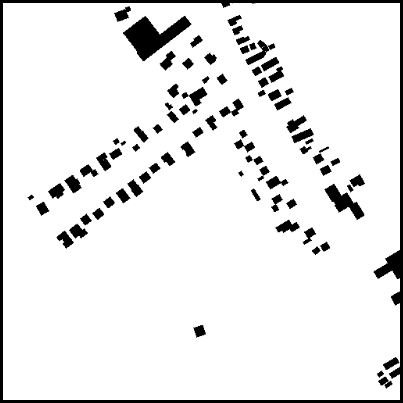}
    \caption{Urban layout}
    \label{fig:imageC}
\end{subfigure}
\hfill
\begin{subfigure}[b]{0.225\linewidth}
    \includegraphics[width=\linewidth]{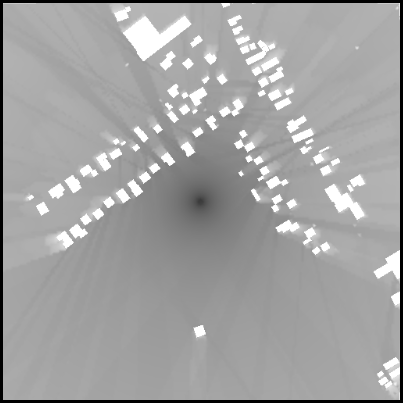}
    \caption{Sound propagation}
    \label{fig:imageD}
\end{subfigure}
\caption{Starting with the selection of a 500m² area (a), buildings are identified, followed by placing a receiver grid (b). The urban layout (c) and the corresponding sound propagation, simulated using the NoiseModelling Framework from a central signal source, are then used in the dataset (d).}
\label{fig:overview}
\end{figure}

\noindent \textbf{Limitation to 2d Propagations.} A significant limitation of our dataset is the absence of building height information, which restricts our ability to accurately model vertical sound diffraction. Consequently, our current dataset focuses only on horizontal diffraction. Vertical diffraction is disabled during the simulation step.

\subsection{Prediction Tasks}

We simulate the \textit{ground-truth} propagation maps at all locations for four different prediction tasks with increasing complexity, resulting in a total of $100k$ samples in the final dataset.

\noindent\textbf{Baseline Task:} In the most simplistic baseline setup, a steady noise source with a level of 95 dB at 500 Hz has been simulated without diffraction and reflection. This allows to test generative models under stable conditions, focusing on changes in the environment around the sound source.\\
\noindent\textbf{Diffraction Task:} To isolate the effect of sound wave diffraction around obstacles this setup is nearly identical to the baseline task, except for enabling horizontal diffraction during the simulation process. This task allows to assess the models' precision in predicting how sound waves bend and spread upon encountering buildings.\\
\noindent\textbf{Reflection Task:} Maintaining the same sound level and frequency, this setting focused on modeling sound reflections off surfaces. The constant power level $L^{j}_{W,i}$ and a standardized absorption coefficient $\alpha_{vert}$ (set to 0.1) were used to simulate paths with multiple reflections.\\
\noindent\textbf{Combined Task:} Incorporating variance in source sound levels (60 to 115 dB) and environmental variables like humidity and temperature, this setting added complexity. The power level $L^{j}_{W}$ and atmospheric absorption coefficient $A^{j}_{atm}$ varied per sample. Reflection and diffraction were included to replicate realistic sound propagation, with each phenomenon adding complexity by altering the sound paths and power levels at receivers.

\section{Evaluation \& Baseline Results}

\subsection{Experimental Setup}
\paragraph{Generative Models.} For a first baseline evaluation, we benchmarked three widely used architectures for image generation on all four tasks: a standard U-Net~\cite{unet}, a Pix2Pix-based GAN~\cite{pix2pix}, and a Denoising Diffusion Probabilistic Model~\cite{ddpm}. Each model was constructed upon a unified U-Net backbone, scaling from 64 to 1028 channels, and then reconverging to 64. All models were trained with a consistent batch size of 18 for a maximum of 50 epochs. An early stopping criterion with a patience of 3 epochs was implemented to curtail training when no improvement in validation loss was detected, thereby preventing overfitting. All additional hyperparameters can be found in Appendix~\ref{hyperparamter}.\\
\noindent\textbf{Evaluation Metrics.}
The primary interest of our baseline evaluation is to investigate the capabilities of standard image-to-image generative models to predict physically correct sound propagations in our defined tasks. The pixel difference at a given location between true and prediction is used to determine the quality of this prediction. While Mean Absolute Error (MAE) quantifies the average magnitude of prediction errors, the implementation of Weighted Mean Absolute Percentage Error (wMAPE) enhances this evaluation by particularly penalizing inaccuracies in instances where predictions gives high values in areas that physically should have low amplitudes, such as regions behind buildings. This approach assigns a maximal error rate of 100\% to these errors. Additionally, to assess how well each model captured the reflections or diffractions we specifically measured both metrics in areas not in direct line of sight to the central signal source in the OSM images. This assessment was conducted by performing ray tracing from the sound source, allowing us to evaluate the models' effectiveness in prediction propagations in Line-of-Sight (LoS) and outside Line-of-Sight (NLoS) by reflection and diffraction.

\subsection{Baseline Performance} 
The initial assessment of U-Net, GAN, and Diffusion models using MAE and MAPE established a baseline for comparison (see Table\ref{tab:performance_metrics}). The baseline provides a controlled environment to assess the generative models' fundamental capabilities. With a fixed sound level and frequency, the models' performance in predicting $L^{j}_{{R_k}}$ could be evaluated without the additional complexities of environmental interactions. This setup confirmed that even under stable conditions, there is some variance in the models ability to capture the subtle changes around the noise source.

\begin{table}[ht]
  \caption{Performance metrics including MAE and MAPE across different tasks for all architectures. Note: Due to varying decibel value ranges, the combined task metrics cannot be directly compared with the other three tasks.}
  \label{tab:performance_metrics}
  \centering
  \begin{tabular}{lcccccccc}
    \textbf{Model} & \multicolumn{2}{c}{\textbf{Baseline}} & \multicolumn{2}{c}{\textbf{Diffraction}} & \multicolumn{2}{c}{\textbf{Reflections}} & \multicolumn{2}{c}{\textbf{Combined}} \\
    & MAE & wMAPE & MAE & wMAPE & MAE & wMAPE & MAE & wMAPE \\
    \hline
    UNet & 2.08 & 19.45 & \textbf{1.65} & 9.75 & 3.22 & 31.87 & 1.77 & 20.59 \\
    GAN & \textbf{1.52} & \textbf{8.21} & 1.66 & \textbf{8.03} & \textbf{2.88} & \textbf{16.57} & 1.76 & \textbf{19.12} \\
    Diffusion & 2.57 & 25.21 & 2.12 & 11.85 & 4.14 & 35.20 & \textbf{1.57} & 21.45 \\
  \end{tabular}
\end{table}

\begin{figure}[h]
\begin{center}
%\framebox[4.0in]{$\;$}
\includegraphics[width=1\linewidth]{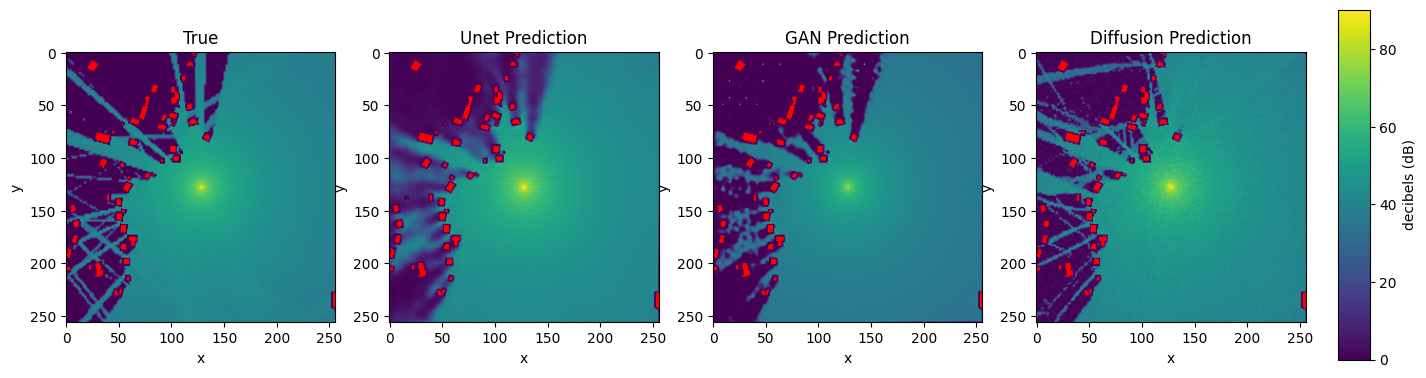}
\end{center}
\caption{Comparing the ground-truth simulation with the predictions from U-Net, GAN, and diffusion model for a single sample within the \textbf{reflection} task.}
\label{fig:ref}
\end{figure}

\noindent\textbf{Reflection and Diffraction Tasks.} 
For the reflection task, each model displayed distinct performance characteristics regarding their ability to model reflective patterns. A sample from the reflection dataset, as illustrated in Figure~\ref{fig:ref}, showcases the unique approaches of each model architecture in handling the higher-order complexities of reflections (see eq. \ref{eq:reflection}). More qualitative samples are provided in Appendix \ref{app:results}, quantitative results are shown in table \ref{tab:combined_metrics}.

\noindent The U-Net model tends to blur most areas, indicating a general approach to sound mapping that prioritizes broad patterns over specific details. On the other hand, the GAN model attempts to imprint specific patterns within these areas, reflecting a more detailed-oriented strategy. Notably, the Diffusion model, despite having a high MAE and MAPE in NLoS areas, makes an apparent effort to visually replicate reflection patterns.

\noindent In the reflection task, the recursive nature of sound interactions (see eq. \ref{eq:reflection}) introduces non-linear complexities. The nonlinearity arises because the adjustment for each reflection depends on the cumulative effect of all previous reflections, each potentially altering the sound level in a unique manner. The generative models must, therefore, be capable not just of modeling the initial interaction of sound with the environment, but also of capturing the compounded effect of multiple reflections.\\
\noindent\textbf{Combined Task.} The final comprehensive evaluation using variance in all terms revealed the adaptability of the Diffusion model (see table \ref{tab:combined_metrics}), showing a marked improvement in its performance. Meanwhile, U-Net and GAN exhibited a stable performance across different tasks, highlighting their consistency.

\begin{figure}[h]
\begin{center}
%\framebox[4.0in]{$\;$}
\includegraphics[width=1\linewidth]{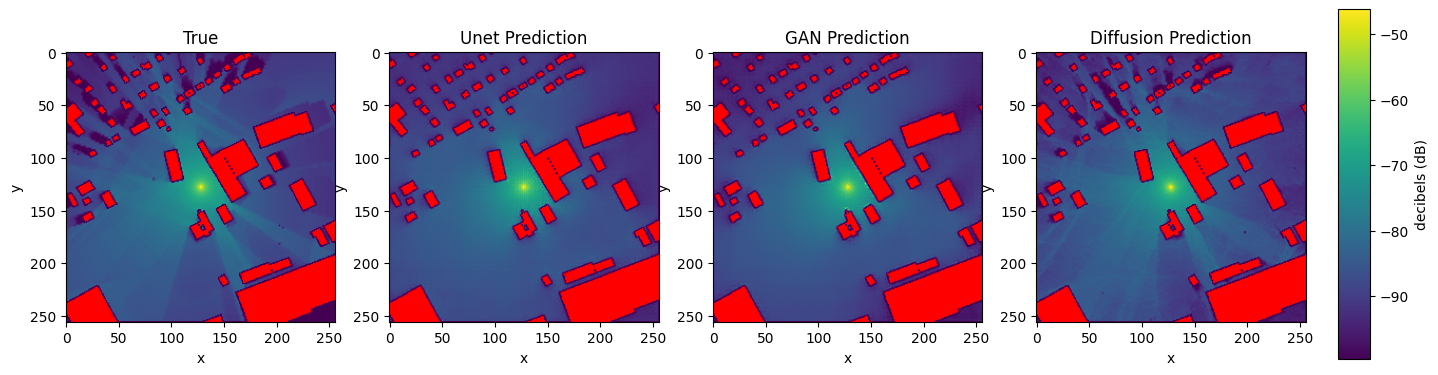}
\end{center}
\caption{Comparing the ground-truth simulation with the prediction of the diffusion model for a single sample within the \textbf{combined task}, distinguishing between the MAE in LoS and NLoS.}
\label{fig:sight_loss}
\end{figure}

\subsection{Analysis}

By separating the loss in LoS and NLoS conditions, each model exhibited distinct strengths and weaknesses. When models encounter tasks with reflections or diffractions, we observe an increase in MAE for NLoS regions across all tested architectures. Both the UNet and GAN models show a moderate rise in NLoS error, but it is most significant in the Diffusion model. Despite its ability to visually replicate sound reflection patterns, there is a notable discrepancy between its visual outputs and actual acoustic precision. This discrepancy suggests that the Diffusion model, while visually detailed, does not fully capture the complexities of sound physics.

\begin{figure}[h]
\begin{center}
%\framebox[4.0in]{$\;$}
\includegraphics[width=1\linewidth]{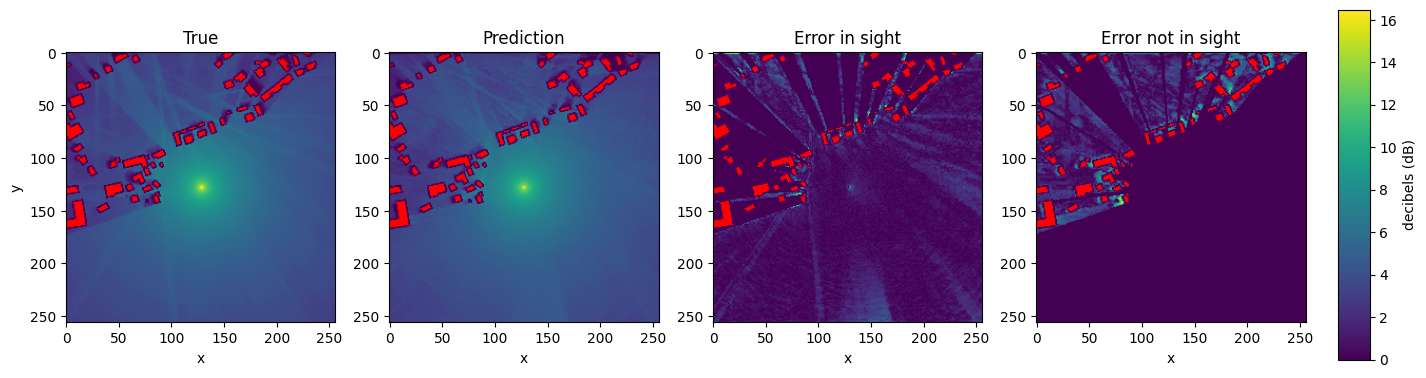}
\end{center}
\caption{Comparing the ground-truth simulation with the prediction of the diffusion model for a single sample within the \textbf{reflection} task, distinguishing between the MAE in LoS and NLoS.}
\label{fig:sight_loss}
\end{figure}

\begin{table}[ht]
  \caption{Consolidated performance metrics across tasks for all architectures. The "Runtime per Sample" is reported in seconds and represents an average computed over 100 samples.}
  \label{tab:combined_metrics}
  \begin{center}
    \begin{tabular}{lllllll}
      \textbf{Model} & \textbf{Task} & \textbf{LoS} & \textbf{NLoS} & \textbf{LoS} & \textbf{NLoS} & \textbf{Runtime per} \\
      & & \textbf{MAE} & \textbf{MAE} & \textbf{wMAPE} & \textbf{wMAPE} & \textbf{Sample (s)}\\
      \hline
      Sim. & Baseline & 0.00 & 0.00 & 0.00 & 0.00 & 20.4717 $\pm{1.4885}$ \\
      UNet & Baseline & 2.29 & 1.73 & 12.91 & 37.57 &  0.0126 $\pm{0.0012}$ \\
      GAN & Baseline & \textbf{1.73} & \textbf{1.19} & \textbf{9.36} & \textbf{6.75} &  \textbf{0.0095 \textpm 0.0012} \\
      Diffusion & Baseline & 2.42 & 3.26 & 15.57 & 51.08 &  4.1560 $\pm{0.0061}$ \\
      \hline
      Sim. & Diffraction & 0.00 & 0.00 & 0.00 & 0.00 & 20.6027 $\pm{0.7953}$ \\
      UNet & Diffraction & 0.94 & \textbf{3.27} & 4.22 & 22.36 & 0.0126 $\pm{0.0012}$ \\
      GAN & Diffraction & \textbf{0.91} & 3.36 & \textbf{3.51} & \textbf{18.06} &  \textbf{0.0095 \textpm 0.0012} \\
      Diffusion & Diffraction & 1.59 & \textbf{3.27} & 8.25 & 20.30 &  4.1560 $\pm{0.0061}$ \\
      \hline
      Sim. & Reflection & 0.00 & 0.00 & 0.00 & 0.00 & 25.0973 $\pm{3.2928}$ \\
      UNet & Reflection & 2.29 & 5.72 & 12.75 & 80.46 &  0.0126  $\pm{0.0012}$ \\
      GAN & Reflection & \textbf{2.14} & \textbf{4.79} & \textbf{11.30} & \textbf{30.67} &  \textbf{0.0095 \textpm 0.0012} \\
      Diffusion & Reflection & 2.74 & 7.93 & 17.85 & 80.38 &  4.1560 $\pm{0.0061}$ \\
      \hline
      Sim. & Combined & 0.00 & 0.00 & 0.00 & 0.00 & 29.2395 $\pm{4.7059}$ \\
      UNet & Combined & 1.39 & 2.63 & 10.10 & 45.15 &  0.0126 $\pm{0.0012}$ \\
      GAN & Combined & 1.37 & 2.67 & \textbf{9.80} & 40.68 &  \textbf{0.0095 \textpm 0.0012} \\
      Diffusion & Combined & \textbf{1.26} & \textbf{2.21} & 13.07 & \textbf{40.38} &  4.1560 $\pm{0.0061}$ \\    \end{tabular}
  \end{center}
\end{table}

\noindent An illustrative example from the final setting incorporating a complex mix of direct sound paths, reflections, and diffractions, with varying environmental conditions is depicted in Figure~\ref{fig:sight_loss}. This sample visualizes the nuanced challenges faced by the diffusion model. More qualitative results are provided in Appendix \ref{app:results}, quantitative results are shown in table \ref{tab:combined_metrics}.
\\

\noindent\textbf{Runtime Analysis}
The comparison of performance metrics in Table \ref{tab:combined_metrics} reveals differences between generative models and the traditional sound propagation simulation in terms of processing speed. The generative models, especially GAN and U-Net, show a significant improvement in runtime over the conventional simulation (up to factor $20k$). The Diffusion model has a slightly higher runtime than GAN and U-Net. This comparison underscores the advantages of GAN and U-Net in terms of speed, highlighting their effectiveness as quicker alternatives to traditional simulation methods. For additional analysis, please refer to Appendix~\ref{tab:model_simulation_runtime_comparison}.
\section{Discussion \& Future Work}
Our first baseline results showed that the proposed dataset provides a suitable proxy-problem for further research and development of data driven models for the prediction of complex physical tasks. The provided 2d sound propagation tasks have a manageable compute complexity, both on the simulation side as well for the training of state-of-the-art generative models, while providing different levels of difficulty. The evaluation of current image-to-image generative models shows, that speedups of up to factor $20k$ compared to the simulation are realistic, while the physical correctness still needs further improvement. \\
The initial analysis of the prediction results points towards two very interesting phenomena, which need further investigation: I) the very different error pattern produced by different generative approaches, and II) the eminent disability of all model to capture higher order dependencies.

\noindent\textbf{Broader Impact Statement.} The aim of the presented dataset is to foster research towards fast and physically correct 1-step generative models. Such models have the potential to drastically speed-up complex simulation of environmental (climate) or engineering problems, resulting in a wide range of positive effects in various applications. As for any simulation technique, the authors can not entirely rule out malignant applications.   
\newpage
\section*{Reproducibility Statement}
\label{sec:reproducibilty_statement}
Detailed information about the dataset and the download link are provided under the website: \url{https://www.urban-sound-data.org/}. The code necessary to replicate the experiments discussed in this paper has been made publicly available on the GitHub website: \url{https://github.com/urban-sound-data/urban-sound-data}. Additionally, a comprehensive description of the dataset is given as a Datacard in Section~\ref{sec:datacard}.

\section*{Funding Acknowledgement}
The authors acknowledge the financial support by the German Federal 
Ministry of Education and Research (BMBF) in the program “Forschung an Fachhochschulen in Kooperation mit Unternehmen (FH-Kooperativ)” within the joint project "KI-Bohrer" under 
grant 13FH525KX1.

%Bibliography
\bibliographystyle{unsrt}  
\bibliography{iclr2024_workshop}
\newpage
\appendix

\section{Scalable Simulation Pipeline.\label{app:pipeline}}
\begin{figure}[h]
\begin{center}
\includegraphics[width=1\linewidth]{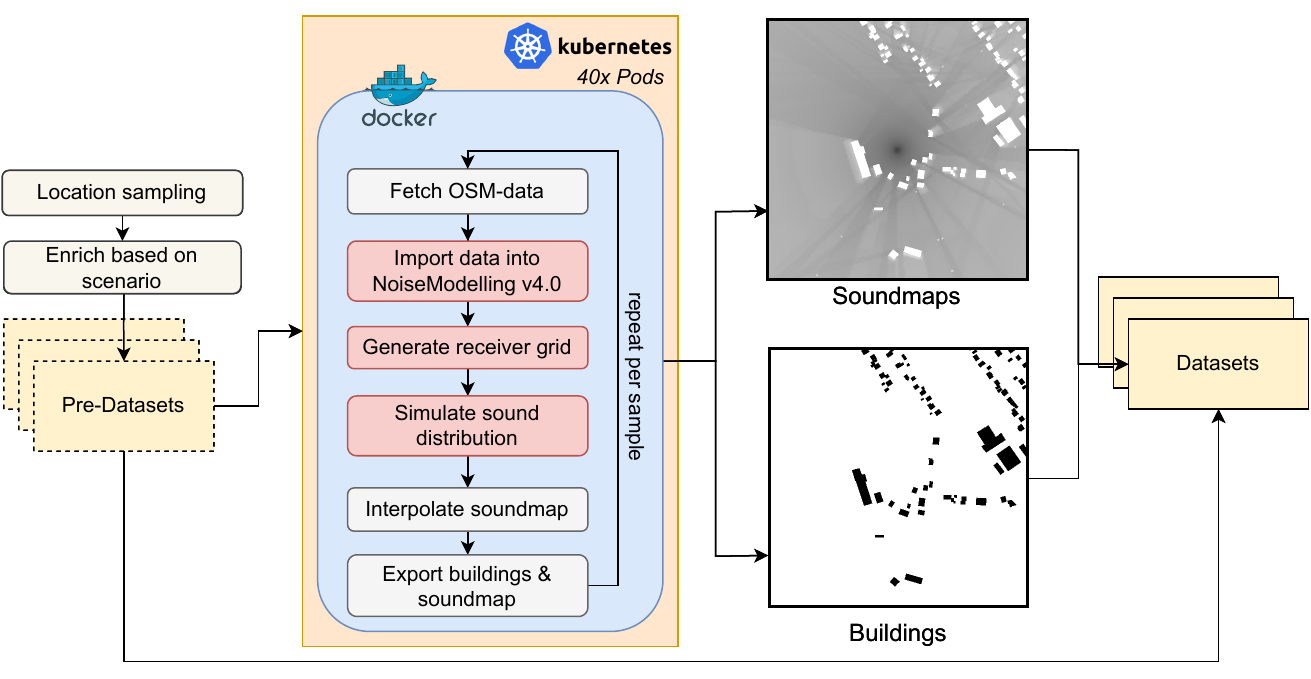}
\end{center}
\caption{Detailed visualization of the dataset generation pipeline.}
\label{fig:pipeline}
\end{figure}

The dataset generation pipeline visualized in Figure~\ref{fig:pipeline} is a crucial component of our study, developed to efficiently process sound propagation data in diverse urban settings. Utilizing the NoiseModelling framework~\cite{noisemodelling_framework}, we have automated the data input and simulation processes within a Docker-containerized environment. The pipeline commences with the automatic download of a 500m² area map from OpenStreetMap for each location using the Overpass API, followed by their import into the NoiseModelling framework alongside the signal source.

\noindent Considering the computational intensity of this process, with an average duration of 30 seconds per sample, our pipeline is structured for scalability. It operates on a Kubernetes cluster with 40 pods, enabling us to complete the generation of the entire dataset, encompassing 25,000 data points for each complexity level, in approximately 20 hours.\\

\section{Additional Qualitative Results \label{app:results}}

\begin{figure}[H]
\centering
\begin{subfigure}[b]{0.9\linewidth}
    \includegraphics[width=\linewidth]{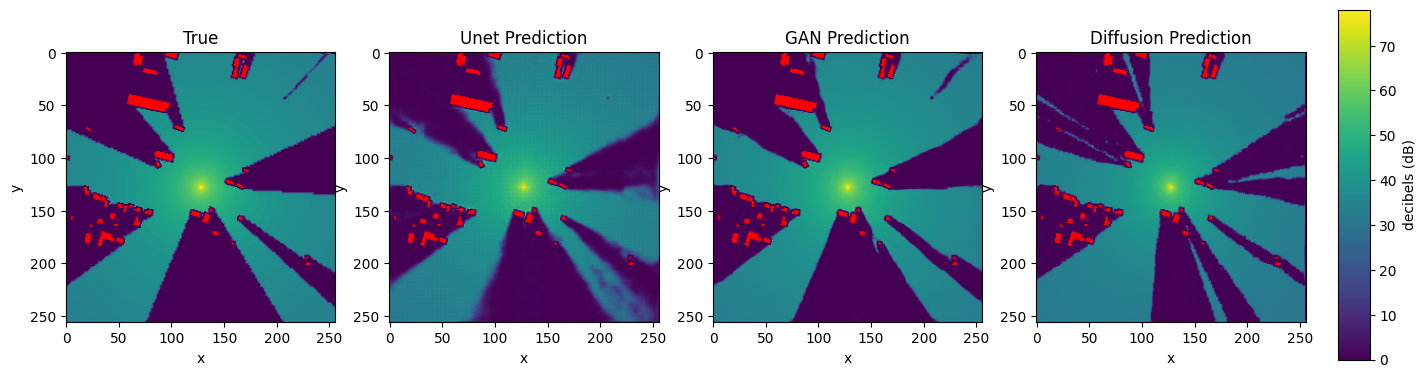}
    \label{fig:imageA}
\end{subfigure}
\vfill
\begin{subfigure}[b]{0.9\linewidth}
    \includegraphics[width=\linewidth]{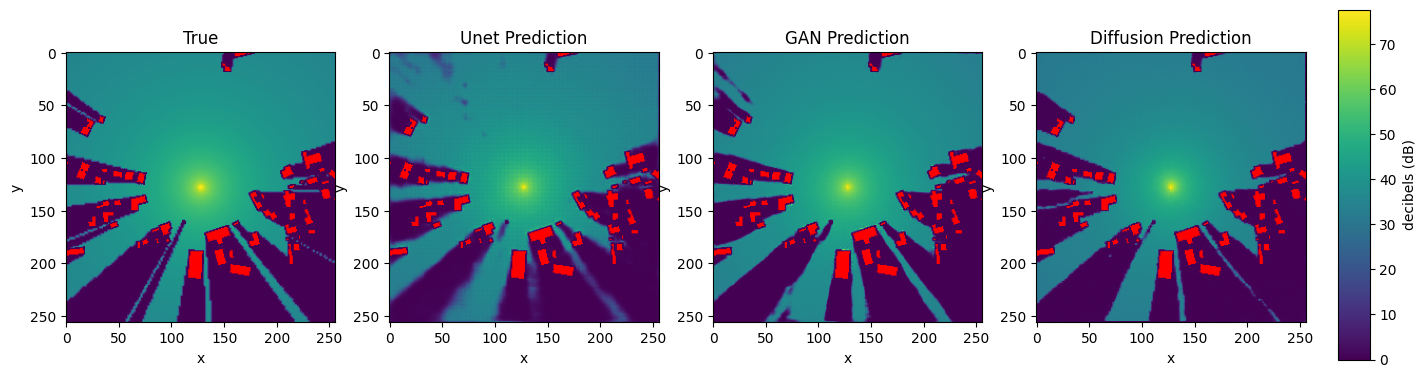}
    \label{fig:imageB}
\end{subfigure}
\caption{Comparing the output of the physical simulation with the predictions from U-Net, GAN, and diffusion model for a single sample within the \textbf{baseline} task dataset.}
\label{fig:overview}
\end{figure}

\begin{figure}[H]
\centering
\begin{subfigure}[b]{0.9\linewidth}
    \includegraphics[width=\linewidth]{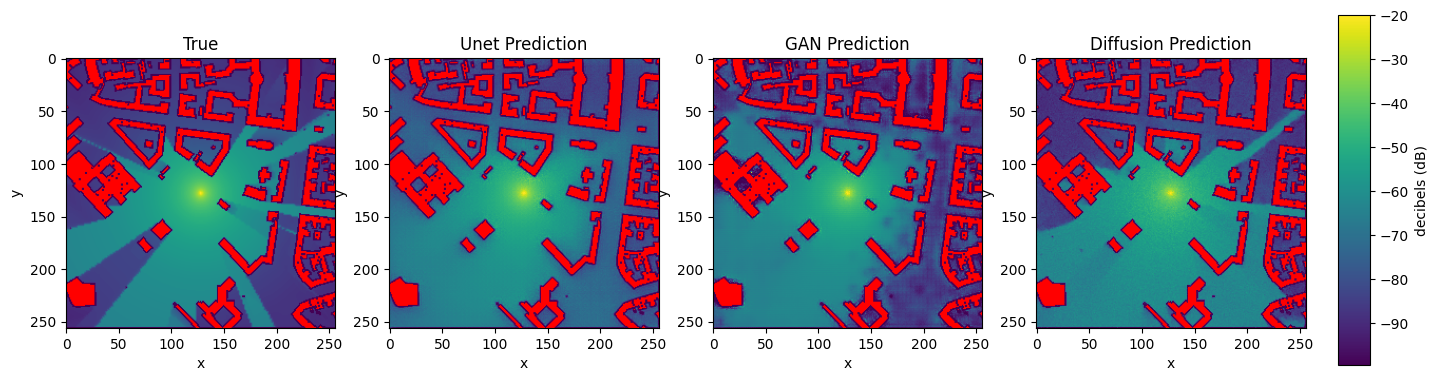}
    \label{fig:imageA}
\end{subfigure}
\vfill
\begin{subfigure}[b]{0.9\linewidth}
    \includegraphics[width=\linewidth]{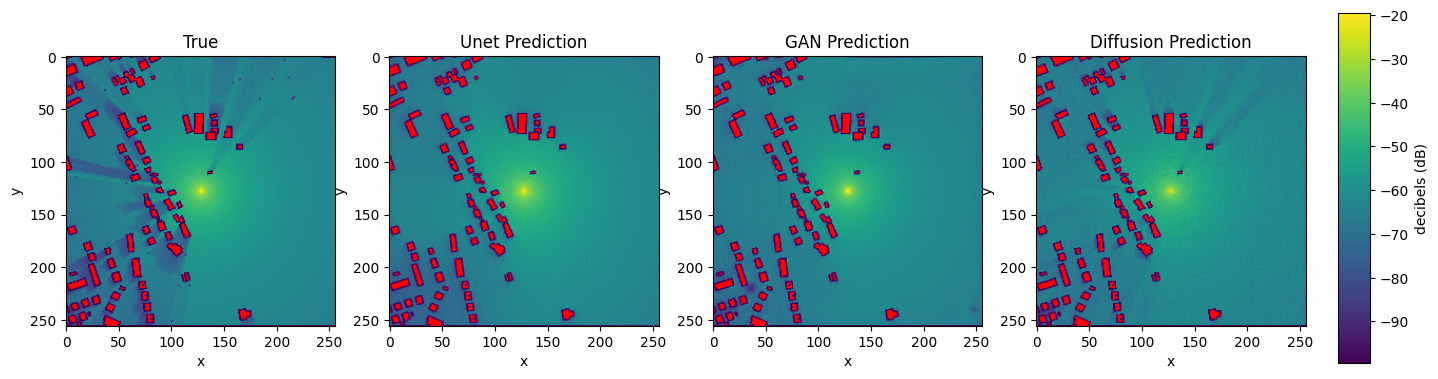}
    \label{fig:imageB}
\end{subfigure}
\caption{Comparing the output of the physical simulation with the predictions from U-Net, GAN, and diffusion model for a single sample within the \textbf{diffraction} task dataset.}
\label{fig:overview}
\end{figure}

\begin{figure}[H]
\centering
\begin{subfigure}[b]{0.9\linewidth}
    \includegraphics[width=\linewidth]{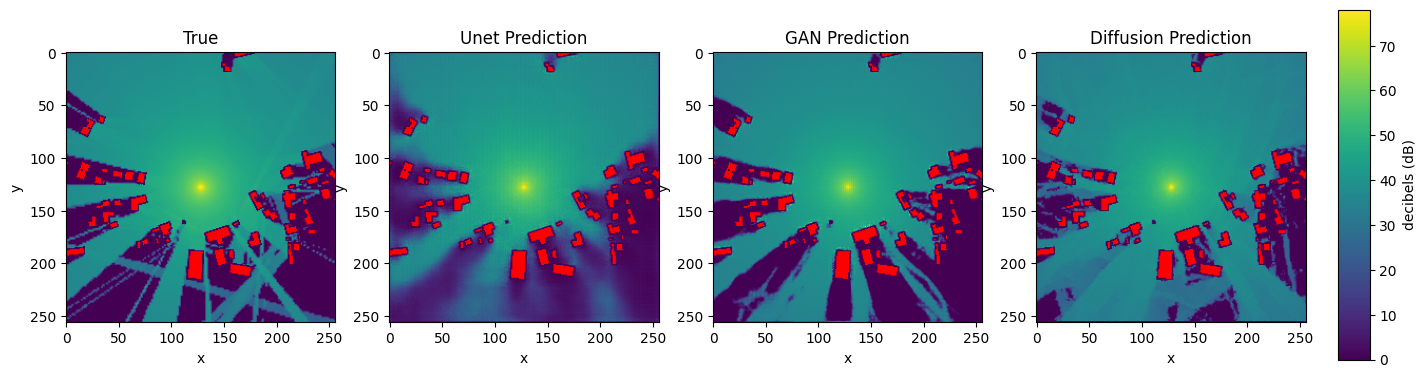}
    \label{fig:imageA}
\end{subfigure}
\vfill
\begin{subfigure}[b]{0.9\linewidth}
    \includegraphics[width=\linewidth]{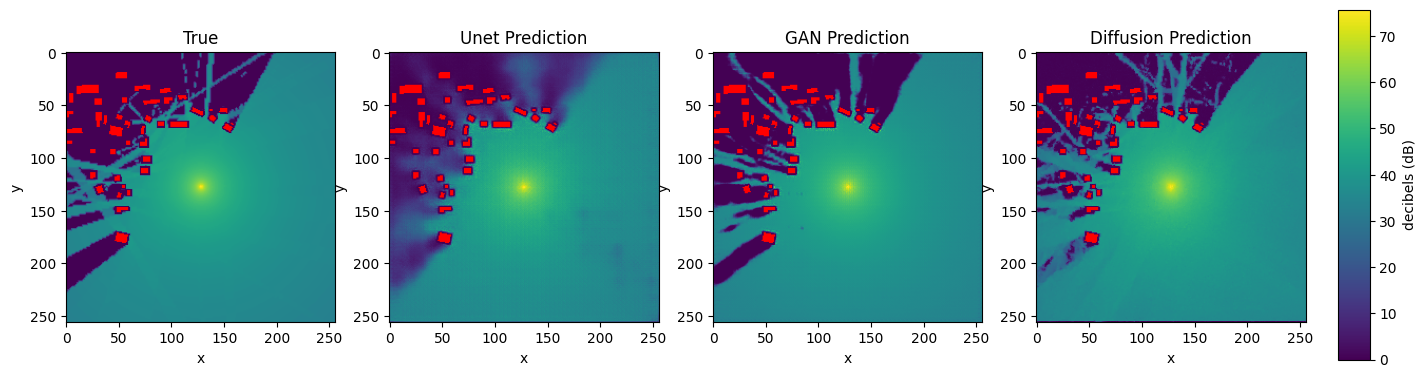}
    \label{fig:imageB}
\end{subfigure}
\caption{Comparing the output of the physical simulation with the predictions from U-Net, GAN, and diffusion model for a single sample within the \textbf{reflection} task dataset.}
\label{fig:overview}
\end{figure}

\begin{figure}[H]
\centering
\begin{subfigure}[b]{0.9\linewidth}
    \includegraphics[width=\linewidth]{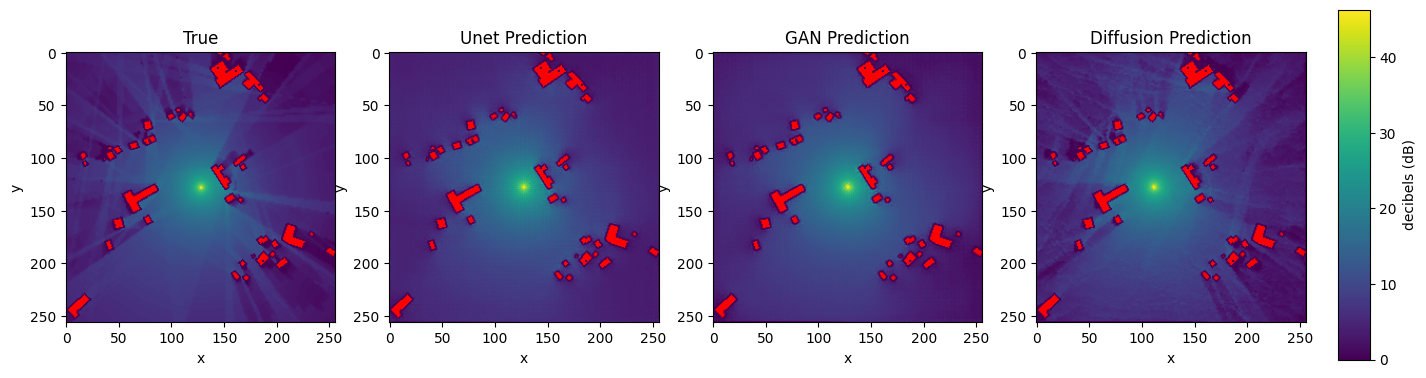}
    \label{fig:imageA}
\end{subfigure}
\vfill
\begin{subfigure}[b]{0.9\linewidth}
    \includegraphics[width=\linewidth]{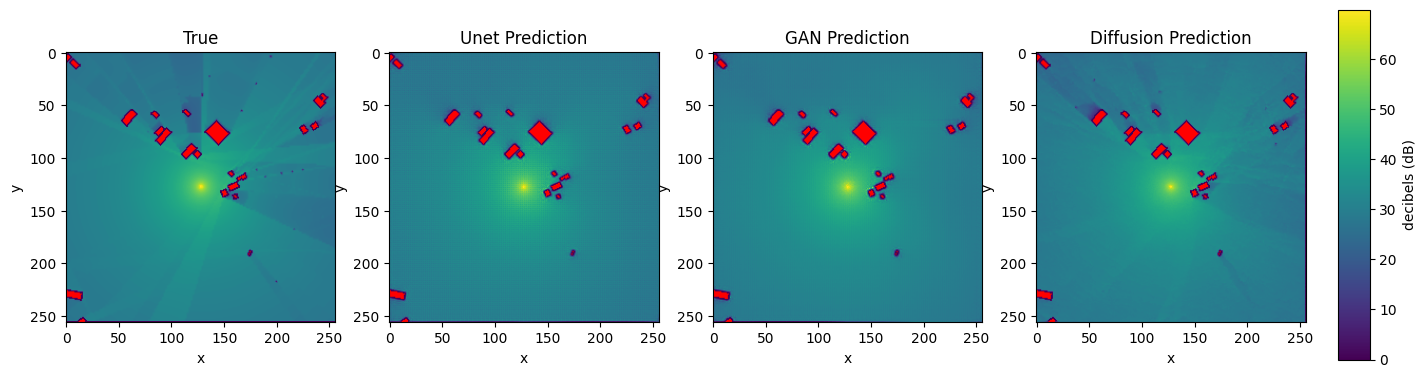}
    \label{fig:imageB}
\end{subfigure}
\caption{Comparing the output of the physical simulation with the predictions from U-Net, GAN, and diffusion model for a single sample within the \textbf{combined} task dataset.}
\label{fig:overview}
\end{figure}

\begin{figure}[H]
\centering
\begin{subfigure}[b]{1\linewidth}
    \includegraphics[width=\linewidth]{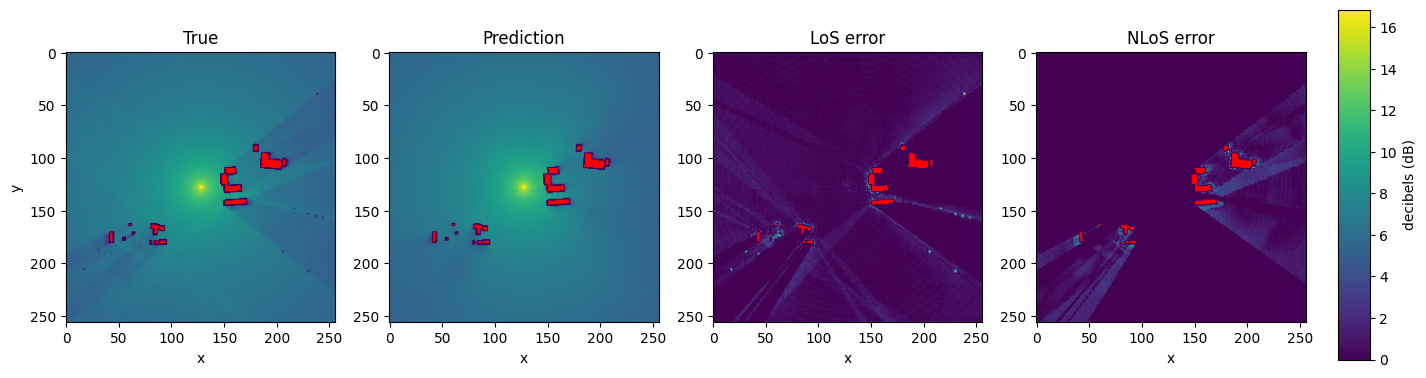}
    \caption{U-Net}
    \label{fig:imageA}
\end{subfigure}
\vfill
\begin{subfigure}[b]{1\linewidth}
    \includegraphics[width=\linewidth]{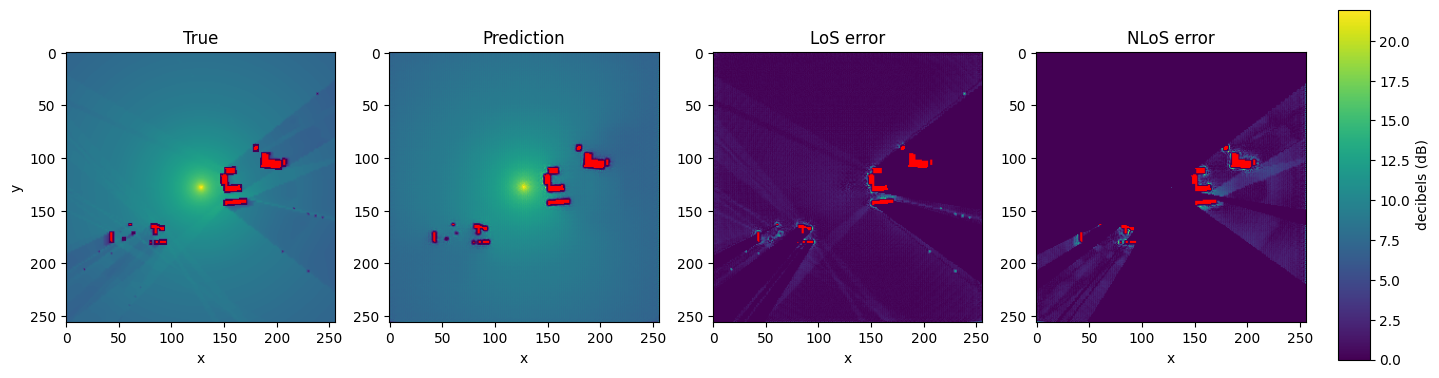}
    \caption{GAN}
    \label{fig:imageB}
\end{subfigure}
\vfill
\begin{subfigure}[b]{1\linewidth}
    \includegraphics[width=\linewidth]{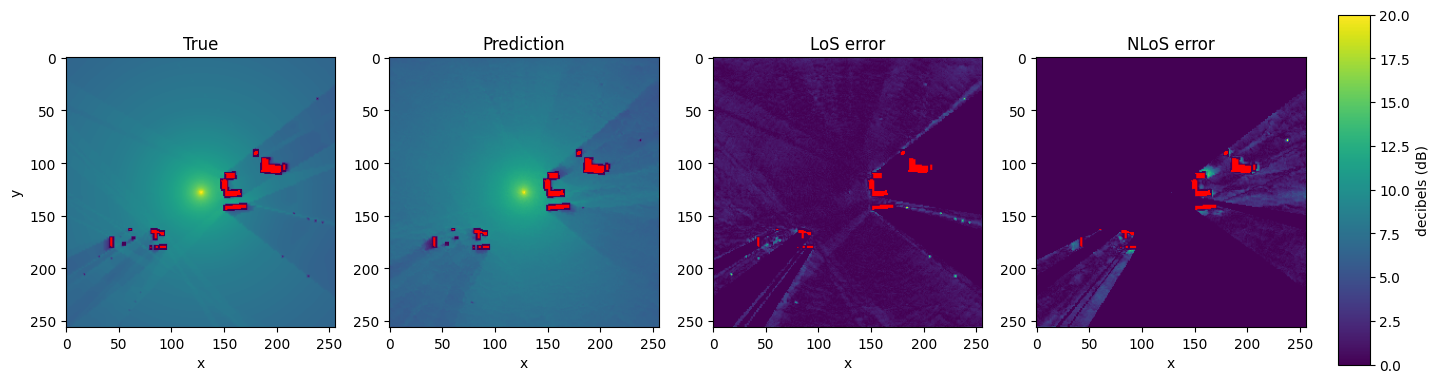}
    \caption{Diffusion model}
    \label{fig:imageC}
\end{subfigure}
\caption{Comparing the output of the physical simulation with the prediction of U-net (a), GAN (b) and diffusion model (c) for a single sample within the \textbf{reflection} task dataset, distinguishing between the MAE in LoS and NLoS.}
\label{fig:overview}
\end{figure}

\section{Training Setup}
\label{hyperparamter}

This appendix provides a detailed overview of the architecture, input specifications and hyperparameter used for the generative models utilized in this research. The U-Net model architecture, adopted from~\cite{unet}, and the GAN setup, based on~\cite{pix2pix}, are designed to process grayscale image inputs. For the \textbf{combined} task, the input is extended by appending additional parameters as a separate dimension.

For the Diffusion model, we followed the methodology described in~\cite{ddpm} while incorporating conditional inputs as separate dimensions besides the noised input image. Each model was constructed upon a unified U-Net backbone, scaling from 64 to 1028 channels, and then reconverging to 64.

\begin{table}[H]
\centering
\caption{UNet Training Hyperparameter}
\label{tab:hyp_unet}
\begin{tabular}{l|l}
\hline
\textbf{Hyperparameter} & \textbf{Value} \\
\hline
Batchsize & 18 \\
Learning Rate & $1 \times 10^{-4}$ \\
Image Size & 256 \\
Max Epochs & 50 \\
\hline
\end{tabular}
\end{table}

\begin{table}[H]
\centering
\caption{GAN Training Hyperparameter}
\label{tab:hyp_gan}
\begin{tabular}{l|l}
\hline
\textbf{Hyperparameter} & \textbf{Value} \\
\hline
Batchsize & 18 \\
Learning Rate Discriminator & $1 \times 10^{-4}$ \\
Learning Rate Generator & $2 \times 10^{-4}$ \\
Image Size & 256 \\
Max Epochs & 50 \\
L1 Lambda & 100 \\
Lambda GP & 10 \\
\hline
\end{tabular}
\end{table}

\begin{table}[H]
\centering
\caption{Diffusion Model Training Hyperparameter}
\label{tab:hyp_dif}
\begin{tabular}{l|l}
\hline
\textbf{Hyperparameter} & \textbf{Value} \\
\hline
Batchsize & 18 \\
Learning Rate & $1 \times 10^{-4}$ \\
Image Size & 256 \\
Max Epochs & 50 \\
Noise Steps & 1000 \\
\hline
\end{tabular}
\end{table}

\begin{table}[H]
  \caption{Model vs. Simulation Performance Comparison for Single Sample Processing}
  \label{tab:model_simulation_runtime_comparison}
  \begin{center}
    \begin{tabular}{lcc}
      \textbf{Model - Condition} & \textbf{Mean Runtime (s)} & \textbf{Std. Dev. (s)} \\
      \hline
      UNet & 0.0126 & 0.0012 \\
      GAN & 0.0095 & 0.0012 \\
      Diffusion & 4.1560 & 0.0061 \\
      \hline
      Simulation - Baseline & 20.4717 & 1.4885 \\
      Simulation - Diffraction & 20.6027 & 0.7953 \\
      Simulation - Reflection & 25.0973 & 3.2928 \\
      Simulation - Combined & 29.2395 & 4.7059 \\
      \\
      Simulation - Combined & 186.2295 & 16.8491 \\
      - 3rd Order Reflections &  &  \\
    \end{tabular}
  \end{center}
\end{table}
\label{sec:datacard}


\section*{Supplementary material}

# Urban Sound Propagation

This dataset is assembled for research into urban sound propagation, comprising 25,000 data points across 10 diverse cities. Each city is represented by 2,500 locations, offering a comprehensive analysis of various urban configurations. The dataset utilizes OpenStreetMap (OSM) imagery to detail the urban layout within a 500m x 500m area for each location, where buildings are delineated with black pixels and open spaces with white pixels.

Supplementing the urban structural images, the dataset includes sound distribution maps at resolutions of 512x512 and 256x256. These maps are precisely generated through the NoiseModelling v4.0 framework, an advanced simulation tool engineered for accurate modeling of sound dynamics within urban environments.

For researchers and experts interested in exploring the intricacies of sound simulation, additional insights can be obtained from the NoiseModelling framework documentation.

| DATASET LINK | DATA CARD AUTHOR(S) |
|---|---|
| https://doi.org/10.5281/zenodo.10609793 | **Martin Spitznagel, IMLA**: Owner<br>**Janis Keuper, IMLA**: Owner |

# Authorship

## Publishers

| PUBLISHING ORGANIZATION(S) | INDUSTRY TYPE(S) | CONTACT DETAIL(S) |
|---|---|---|
| Institute for Machine Learning and Analytics (IMLA) | Academic – Tech | **Affiliation:** Offenburg University<br>**Website:** https://imla.hs-offenburg.de/ |

## Funding Sources

| INSTITUTION(S) | FUNDING OR GRANT SUMMARY(IES) |
|---|---|
| German Federal Ministry of Education and Research (BMBF) | This project is part of the "Forschung an Fachhochschulen in Kooperation mit Unternehmen (FH-Kooperativ)" program, within the joint project KI-Bohrer [https://www.ki-bohrer.de/] and is funded under the grant number 13FH525KX1. |

# Dataset Overview

| DATA SUBJECT(S) | DATASET SNAPSHOT | CONTENT DESCRIPTION |
|---|---|---|
| Data about natural phenomena<br><br>Data about places and objects | <table><tr><td>Size of Dataset</td><td>~5 GB</td></tr><tr><td>Number of Instances</td><td>~100,000</td></tr><tr><td>Training</td><td>19908 x 4</td></tr><tr><td>Evaluation</td><td>3732 x 4</td></tr><tr><td>Test</td><td>1244 x 4</td></tr></table><br>**Additional Notes:** The dataset is segmented into four distinct subsets, each tailored to explore specific aspects of sound propagation in urban environments: Baseline, Reflection, Diffraction, and Combined. | A data point in this dataset consists of two main components: an urban layout image from OpenStreetMap and a corresponding sound distribution map. The urban layout image is a 500m x 500m area depiction where buildings are marked in black and open spaces in white. The sound distribution map generated using NoiseModelling v4.0, illustrates the sound dynamics within that urban environment at resolutions of 512x512 or 256x256.<br><br>The dataset comprises four subsets: **Baseline** for basic sound behavior, **Reflection** examining sound wave interactions with surfaces, **Diffraction** focusing on sound navigating around objects, and **Combined** which merges reflection, diffraction and changing environmental factors like temperature and humidity. |

# Dataset Version and Maintenance

| MAINTENANCE STATUS | VERSION DETAILS | MAINTENANCE PLAN |
|---|---|---|
| Regularly Updated<br><br>(New versions of the dataset have been or will continue to be made available.) | **Current Version:** 2.0<br><br>**Last Updated:** 02/2024<br><br>**Release Date:** 02/2024 | **Feedback:**<br>martin.spitznagel@hs-offenburg.de |

# Example of Data Points

| PRIMARY DATA MODALITY | SAMPLING OF DATA POINTS | DATA FIELDS |
|---|---|---|
| Multimodal<br>- Image Data<br>- Geospatial Data<br>- Tabular Data | | |

| Field Name | Field Value | Description |
|---|---|---|
| lat | float | Latitude of the sound measurement location. |
| long | float | Longitude of the sound measurement location. |
| db | Object | Key-value pairs of sound levels in decibels for a given frequency (lwd{fqz}). |
| soundmap | string | Path to 256x256 resolution sound distribution image. |
| soundmap_512 | string | Path to 512x512 resolution sound distribution image. |
| osm | string | Path to Open Street Map image showing urban layout. |
| temperature | float | Temperature (°C) at the location. |
| humidity | float | Humidity (%) at the location. |
| yaw | float | Orientation of the noise source. Can be empty. |
| sample_id | int | Unique identifier for the data point. |

## TYPICAL DATA POINT

```
{
  "lat": 48.030229082138526,
  "long": 11.367773397906852,
  "db": {"lwd500": 69},
  "soundmap":
  "./soundmaps/256/0_LEQ_256.png",
  "soundmap_512":
  "./soundmaps/512/0_LEQ_512.png",
  "osm":
"./buildings/osm_23747.png",
  "temperature": 12,
  "humidity": 35,
  "yaw": None,
  "sample_id": "23747"
} ```
```

## EXAMPLE OF DATA POINT

Below is an example of an OSM and Simulated Sound Propagation pair:

OSM:

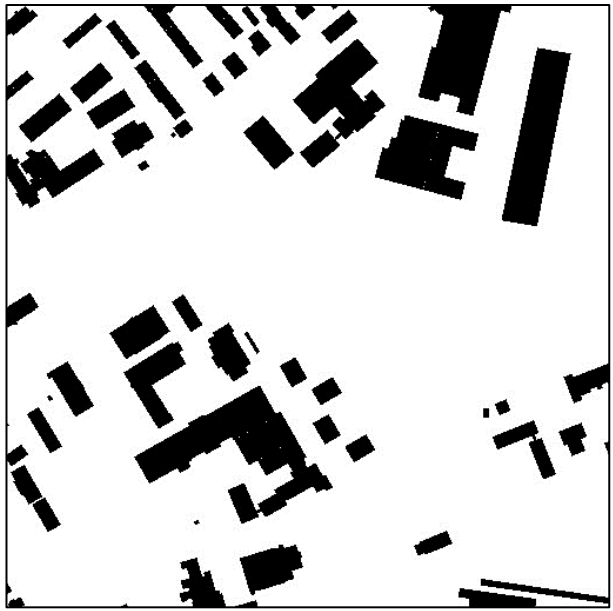

Simulated Sound Propagation:

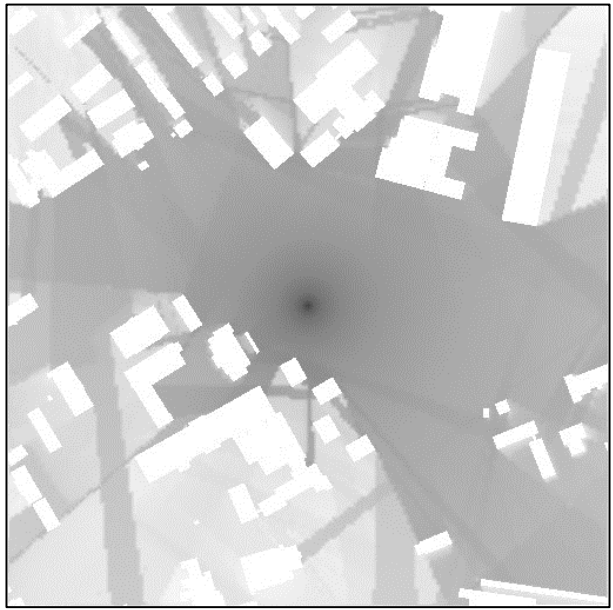

# Motivations & Intentions

## Motivations

| PURPOSE(S) | DOMAIN(S) OF APPLICATION | MOTIVATING FACTOR(S) |
|---|---|---|
| Research | `Generative Models`, `1-step Physic Simulation`, `Sound Propagation`, `Machine Learning` | Generative models, through their capacity to learn from complex datasets, hold significant potential in understanding the intricate physics behind sound propagation. By training on data that encompasses various urban layouts and the corresponding sound distribution maps, these models can uncover the underlying patterns and principles governing how sound travels and interacts with different obstacles, such as buildings and open spaces. This capability enables the creation of predictive models that can simulate sound behavior in any urban environment, thereby offering valuable insights for urban planning, acoustic design, and noise mitigation strategies, all rooted in a deep understanding of the physical laws of sound propagation. |

## Intended Use

| DATASET USE(S) | SUITABLE USE CASE(S) | UNSUITABLE USE CASE(S) |
|---|---|---|
| Safe for research use | **Sound propagation**: Enhancing models for predicting how sound travels in densely built areas. | **Predicting indoor noise levels**: The dataset is designed for outdoor urban sound distribution, not for indoor environments.<br><br>**Traffic flow or congestion analysis:** It focuses on sound distribution and ignores vehicle movements or traffic patterns. |
|  | RESEARCH AND PROBLEM SPACE(S) |  |
|  | The dataset specifically addresses the problem space of outdoor urban noise propagation. It is intended for developing models that can predict sound distribution around buildings within urban environments. |  |

# Provenance

## Collection

| METHOD(S) USED | METHODOLOGY DETAIL(S) | SOURCE DESCRIPTION(S) |
|---|---|---|
| - API<br>- Physical Simulation Framework | **Overpass API**<br><br>**Source:** The Overpass API is a read-only API that serves up custom selected parts of the OSM map data. It acts as a database over the web: the client sends a query to the API and gets back the data set that corresponds to the query.<br><br>**Platform:** https://overpass-api.de/<br><br>**Is this source considered sensitive or high-risk?** [Yes / **No**]<br><br>**Dates of Collection:** [10 2023 - 12 2024]<br><br>**Primary modality of collected data:**<br>Geospatial Data<br><br>**NoiseModelling v4.0**<br><br>**Source:** An advanced simulation tool engineered for accurate modeling of sound dynamics within urban environments.<br><br>**Platform:** https://github.com/Universite-Gustave-Eiffel/NoiseModelling<br><br>**Is this source considered sensitive or high-risk?** [Yes / **No**]<br><br>**Dates of Collection:** [10 2023 - 12 2024]<br><br>**Primary modality of collected data:**<br>Geospatial Data | **OSM Buildings:** This source provides images from Open Street Map (OSM) that depict urban layouts, specifically focusing on buildings within cities. In these images, black pixels represent buildings, and white pixels indicate open spaces.<br><br>**Sound Propagation:** This component of the dataset involves simulated sound distribution images around urban centers, where the noise source is placed at the center. |

## Collection Criteria

| DATA SELECTION | DATA INCLUSION | DATA EXCLUSION |
|---|---|---|
| **Location Sampling:** The locations are randomly sampled across 10 cities/areas: ["Hamburg", "Hannover", "Augsburg", "Bonn", "Muenchen", "Schwerin", "Berlin", "Paris", "Stuttgart", "Aachen"] | **Enough Obstacles:** At least 10 Buildings within a circle r=200m around the sound source.<br>**No Obstacle to close:** No Building within a r=50m circle around the sound source. | **Additional Notes:** If the Data Inclusion criteria is not met, the data is excluded.<br>No additional exclusion criteria are introduced. |

**Additional Notes:**

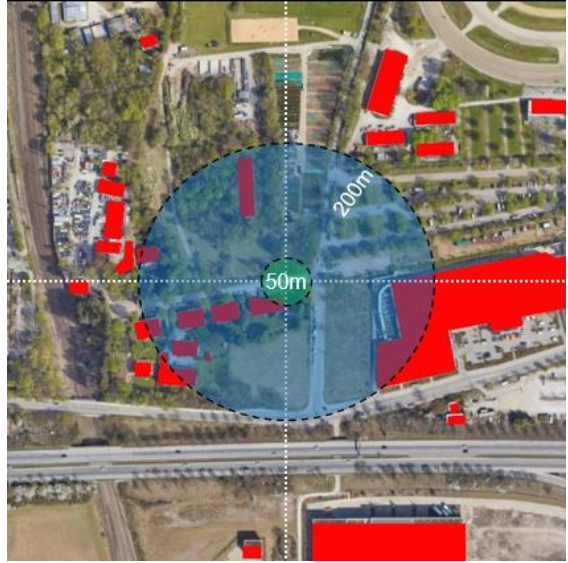

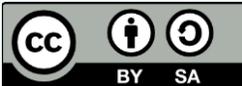



\end{document}